\documentclass[twocolumn]{jpsj2} 
\title{Solid $^4$He and the Supersolid Phase:\\
from Theoretical Speculation to the Discovery of a New State of Matter?\\
A Review of the Past and Present Status of Research}

\author{Davide Emilio \textsc{Galli}\thanks{E-mail address: Davide.Galli@unimi.it}
and Luciano \textsc{Reatto}\thanks{E-mail address: Luciano.Reatto@unimi.it}}
\inst{Dipartimento di Fisica, Universit\'a degli Studi di Milano, Via Celoria 16, 20133 Milano, Italy\\}

\abst{
The possibility of a supersolid state of matter, i.e., a crystalline solid exhibiting 
superfluid properties, first appeared in theoretical studies about forty years ago.
After a long period of little interest due to the lack of experimental evidence, it has 
attracted strong experimental and theoretical attention in the last few years since
Kim and Chan (Penn State, USA) reported evidence for nonclassical 
rotational inertia effects, a typical signature of superfluidity, in samples of solid $^4$He.
Since this ``first observation'', other 
experimental groups have observed such effects in the response to the rotation of samples of
crystalline helium, and it has become clear
that the response of the solid is extremely sensitive to growth conditions, 
annealing processes, and $^3$He impurities.
A peak in the specific heat in the same 
range of temperatures has been reported as
well as anomalies in the elastic behaviour of solid $^4$He with a strong resemblance to the
phenomena revealed by torsional oscillator experiments. Very recently, the observation
of unusual mass transport in hcp solid $^4$He has also been reported, suggesting superflow.
From the theoretical point of view, powerful simulation methods have been used to study
solid $^4$He, but the interpretation of the data is still rather 
difficult; dealing with the question of supersolidity means that one has to face not only
the problem of the coexistence of quantum coherence phenomena and crystalline order, 
exploring the realm of spontaneous symmetry breaking and quantum field theory, but
also the problem of the role of disorder, i.e., how defects, such as vacancies, 
impurities, dislocations, and grain boundaries, participate in the
phase transition mechanism. 
}
\kword{supersolid, quantum solid, solid $^4$He, nonclassical rotational inertia,
superfluidity, Bose-Einstein condensation, 
off-diagonal long-range order, zero-point motion, disorder, vacancies
zero-point vacancies, interstitials, vacancy-interstitial pairs, dislocations, grain boundaries}
\begin{document}
\maketitle

\section{Introduction} 

Should we see atoms flowing without resistance in and out of a solid made of the same atoms,
this would certainly be counterintuitive because the same entities have
to simultaneously give the rigidity intrinsic to a solid and provide the moving particles.
Note the difference between this system and the case of a
superconductor, where the flowing charges are the electrons 
and the solidity of the system is given by the ions.
Still, this possibility is what should happen in a supersolid, a state of matter conjectured to exist almost
forty years ago\cite{andreev,chester}, and solid $^4$He appeared to be the most likely candidate.
A plethora of counterintuitive behaviours have their origin in quantum mechanics and in the indistinguishability
of identical particles, and supersolidity also has its roots in quantum mechanics.
In the case of identical particles with integer spin, such as $^4$He, which has zero total spin, 
the extensive occupation of a single quantum state, i.e.,
the phenomenon of Bose-Einstein condensation (BEC),
is at the heart of (in the sense that BEC is proven to imply\cite{leggett1}) superfluidity.
BEC has been recognized in the superfluid phase of liquid $^4$He, and more recently
it has been directly observed in the metastable phases of low-density ultracold trapped atomic systems.
Typical characteristics of the superflow in such phases\cite{leggett2} are nonclassical rotational
inertia (NCRI) and also the absence of dissipation for flow velocities under the critical velocity as well as
the presence of vortices with a velocity flow field with quantized circulation
in units of $h/m$, where $h$ is Planck's constant and $m$ is the mass of the particles.

As already mentioned, a solid phase with such superflow properties seems paradoxical at first sight.
However, from a theoretical point of view, nothing prohibits such an occurrence. In fact,
model quantum systems exist in which the existence of a supersolid phase has been proved
(an example inherent to the topic of this review is presented in ref. \citeonline{reatto})
or found (for quantum lattice models see, for example, 
refs. \citeonline{lattice1,lattice2,lattice3}).
A supersolid phase would correspond to a phase in which two kinds of order are simultaneously present:
crystalline long-range order and off-diagonal long-range order\cite{penrose} (ODLRO).
Thus, in such a state, both order in real space (crystalline order)
and order in momentum space (corresponding to the off-diagonal order) should be simultaneously present.
These two different kinds of order originate from
two different broken symmetries: the translational symmetry, as observed
in any crystalline solid (directly detectable
in the Bragg scattering), and the gauge symmetry, whose breakage allows for the phase coherence
throughout the system.
Our discussion is devoted to solid $^4$He, in which the translational symmetry is spontaneously broken
and only the interatomic interaction plays a role in determining the reference lattice.
We are not going to discuss the cases of
adsorbed phases of helium, of bulk systems of different species such as hydrogen,
of quantum gases in optical lattices, or of lattice models.
Note that lattice models are qualitatively different from the case of $^4$He, because in such models
the Hamiltonian is invariant only for lattice translations so that the spatial broken symmetry
refers to a discrete group, not to a continuous group such as in a quantum solid.

In this review we have not tried to produce an exhaustive
report of the research contributions to this (presently hot) topic, which are
rapidly growing in number. We suggest that the reader also refers to other
reviews\cite{meisel,prokofev1,balibar1} to obtain a more comprehensive
view of what has been achieved and what remains to be achieved in the theoretical and experimental investigation
of a possible supersolid $^4$He phase.

After this brief introduction, the review is structured as follows:
in $\S2$ we will review the main theoretical discussions and some experimental studies 
that appeared
before the torsional oscillator (TO) ``revolution''\cite{chan1,chan2}; in $\S3$ we will discuss
the main experimental results in the last four years;
section 4 is devoted to a discussion of
the theoretical contributions (we will discuss mostly quantum simulation results) 
stimulated by the new experimental findings, whereas $\S5$ contains a discussion of
what remains to be investigated and clariied on the basis of the present status of the research in this field.

\section{The Supersolid Phase before the TO ``Revolution''}

\subsection{The theoretical speculation}
ODLRO was discussed for the first time by Penrose and Onsager\cite{penrose}
when they successfully generalized the notion of BEC for an interacting system of particles.
The one-body density matrix $\rho_1({\vec r}, {\vec r}')$
represents the probability amplitude of destroying a particle in ${\vec r}'$ and creating it in ${\vec r}$,
and ODLRO is defined by a nonzero value of $\rho_1$ in the limit $|{\vec r}-{\vec r}'| \to \infty$.
This implies a macroscopic occupation of a single momentum state, i.e., BEC, because the Fourier transform
of $\rho_1$ represents the momentum distribution.
In the same paper,
by assuming a model for a solid in which particles are localized at their lattice sites
(thus neglecting the possibility of exchanges and then explicitly breaking the Bose symmetry),
they excluded the possibility of the simultaneous presence of crystalline order and ODLRO.
This was not received as a definitive conclusion, and
some years later the theoretical discussion concerned
models for solids in which the assumption of the localization of the particles was weakened.
The failure of the argument of Penrose and Onsager
is only possible in a so-called quantum solid, i.e., a solid
in which the mean-square displacement of the atoms is comparable to the interatomic 
distance. In a quantum solid the standard harmonic approximation
completely breaks down, and it is no longer possible to associate an atom with a specific 
lattice position so that, even in the presence of a translational broken symmetry,
the atoms are substantially delocalized and microscopic processes such as exchange, the formation of
vacancy-interstitial pairs, and more complex processes are allowed.

Because of their low mass and weak interatomic interaction, $^4$He atoms
are very difficult to solidify and, even
at the lowest temperatures, they only solidify if a substantial pressure is imposed (about 25 bar).
Moreover, near the melting density the mean-square oscillation of a $^4$He atom around the equilibrium 
position is a very large fraction of the lattice parameter, approximately 25\%.
Thus, solid $^4$He is the prototype of a quantum solid, and at the 
end of the 1960s it was readily recognized as the most likely candidate for
observing a supersolid phase in nature.
The mechanism for maintaining the atom delocalized and thus indistinguishable
was proposed to be based on the presence of vacancies in the ground state of the solid phase:
the so-called zero-point vacancies.
This would mean that in the ground state
the probability of occupation of a unit cell of the crystal is less than unity or,
more precisely, two in the case of $^4$He, whic solidifies as a hcp
crystal at low pressure.
Often a crystal in
such a state is called an incommensurate crystal, in contrast with a commensurate crystal in which there
is an integer number of atoms in each unit cell of the crystal. If the atoms are bosons, then a
vacancy, i.e., the absence of an atom, also obeys the same statistics, so that at low temperatures
BEC can be expected with the presence of some form of superfluidity.
All solids at a finite temperature contain a finite
concentration of point defects such as vacancies, but this concentration in a classical solid
vanishes exponentially fast as a function of 1/T, the inverse absolute temperature, as T approaches zero.

Andreev and Lifshitz\cite{andreev} suggested that vacancies, being mobile defects in solid $^4$He,
behave as waves in the crystal, and have
an allowed band of energies that correspond to different wavelengths (quasi-momenta)
of these \textit{defectons}; even if a localized vacancy costs a finite energy,
the lower part of the band might be below the energy of the commensurate crystal.
This happens if the width of the vacancy band is sufficently large, and this implies that the tunneling
probability of an atom to the vacancy site is high.
In this case, the ground state develops a finite concentration of zero-point
vacancies, which can undergo BEC.
This picture is based on the assumption of noninteracting vacancies, and Andreev and Lifshitz considered
vacancy-vacancy interactions simply to limit their number. As discussed below,
the assumption of noninteracting vacancies appears
to be incorrect, because even a few vacancies display significant mutual interaction.

Chester's argument\cite{chester} in favour of a supersolid phase of $^4$He was based on 
the demonstration by Reatto\cite{reatto} that a certain class of wave functions, the Jastrow wave functions,
always possess ODLRO, i.e., a finite condensate fraction.
A Jastrow wave function for a system of $\cal N$ particles in volume $\cal V$
is given by the following product of pair functions,
\begin{equation}
\Psi_J (R)= \prod_{i<j}^{\cal N} e^{-{1 \over 2} u(r_{ij})}/Q_J^{1/2} \quad ,
\label{eq1}
\end{equation}
where $r_{ij}=|{\vec r}_i - {\vec r}_j|$ is the distance between particles $i$ and $j$,
$R=\{{\vec r}_1,..,{\vec r}_{\cal N}\}$ is the many-body configuration,
and $Q_J$ is the normalization constant
of $\Psi_J^2$, i.e.,
\begin{equation}
Q_J=\int_{\cal V} dR \prod_{i<j}^{\cal N}e^{-u(r_{ij})}\quad .
\label{eq2}
\end{equation}
The function $u(r)$, usually called the pseudopotential, in the description
of $^4$He systems typically takes the McMillan form, $u(r)=(b/r)^m$, where $b$ and $m$ are positive
variational parameters.
Such a simple form can be used to give a semiquantitative description of liquid $^4$He and, if the value of
the ``core'' parameter $b$ is sufficently large, $\Psi_J$ describes a system in the solid phase\cite{mcmillan}.
Such a class of wave functions has a configurational probability, i.e., the 
square of the modulus of the wave function, which is equal to the configurational distribution function
of a classical system at a fictitious temperature $T^{\star}$ interacting via a two-body potential $v^{\star}(r)$
proportional to the pseudopotential of $\Psi_J$, i.e., $v^{\star}(r)=k_B T^{\star} u(r)$. 
As for any classical solid at a finite temperature, at a density where the system is solid
such a configurational probability must describe a solid with a finite concentration of vacancies\cite{chester};
thus, for this model, zero-point vacancies are present and BEC is already proved\cite{reatto}.
Some years ago, the concentration of vacancies in the state described by a 
Jastrow wave function was estimated\cite{stillinger} to be
about $6\times10^{-6}$;
however, one should keep in mind that $\Psi_J$ gives a poor description of $^4$He in the solid phase,
for instance, the mean-square deviation of an atom from the equilibrium position is about 1/3 of the correct value.
A substantial improvement can be obtained by multiplying the Jastrow wave function by the product
of one-body terms,
usually Gaussians, which localize the atoms around equilibrium lattice positions,
$\{\vec{R}^l\}_j$, explicitly breaking the translational symmetry;
this is the Jastrow-Nosanow (JN) wave function:
\begin{equation}
\Psi_{JN} (R)= \Psi_J (R) \times \prod_{i}^{\cal N} e^{-C|{\vec r}_i-{\vec R}^l_i|^2}/Q_{JN}^{1/2}
\quad ,
\label{eq3}
\end{equation}
where $Q_{JN}$ is the respective normalization.

Following the works by Andreev and Lifshitz and by Chester,
the presence of a supersolid phase was immediately related to the
possibility of measuring nonclassical rotational inertia by
Leggett\cite{leggett3}; a rotating solid
displays an inertia lower than that associated with the rigid rotation of all its atoms
because a fraction of them do not follow the rotation of the container.
Thus, the superfluid fraction is given by
$\rho_s/\rho = (1-I/I_{cl})$, where $I_{cl}$ is the classical moment of inertia, and
the actual moment of inertia,
\begin{equation}
I=\lim_{\omega \to 0}{\partial^2 F \over \partial \omega^2} \quad ,
\label{eq4}
\end{equation}
is expressed as the limit of vanishing angular velocity $\omega$ of the susceptibility of the free energy
to the rotation.
A rigorous upper bound for the superfluid fraction was derived; it showed that at zero temperature the superfluid
fraction is reduced to a value less than 1 when the ground-state wave function lacks
translational invariance and becomes zero in the absence of ``exchanges'' of particles
between lattice sites.
Here a simple comment is worth mentioning because some successive 
contributions to the literature contained a misleading interpretation:
being Leggett's discussion centered on an upper bound, the presence
of ``exchanges'' does not provide a sufficient condition for the observation of NCRI.
Leggett's letter also contained the suggestion of measuring NCRI by an experiment similar to 
the TO experiments which has been realized
in 2004 and thereafter: the dawning of the revolution.

\subsection{Lack of success in early experiments}
The interest in the possibility of a supersolid state in $^4$He
first motivated a number of theoreticians and then some
physicists specialized in low-temperature experiments
to study and search for quantum effects in solid $^4$He.
The research results obtained up to the beginning of the 1990s are reported in
the review of Meisel\cite{meisel}.
During this period experiments did not reveal any signature of such a state, with the exception of
some anomalies in ultrasound measurements\cite{goodkind1}. This led to a
progressive loss of interest in this topic, and the supersolid state faded away as a theoretical dream.
On the theoretical side, the approaches utilized in the 1970s were based either on very simplified models
or on phenomenological models; thus, it was problematic to judge the relevance of their conclusions
for solid $^4$He.
The majority of these studies used quantum lattice models to investigate the properties of
the ground state and the excited states of quantum solids\cite{guyer,mullin,cheng1,cheng2,fisher,imry}.
Other approaches were based on the development of the procedure suggested by Leggett\cite{leggett3} to obtain
improved numerical estimates of an upper bound to the superfluid fraction in the ground state
of solid $^4$He; these calculations used a
phenomenological Gaussian model for the one-particle local density\cite{fernandez,saslow1,saslow2}.
Some generalizations of the two-fluid model were also attempted\cite{saslow2,liu}.
In the absence of experimental evidence for the supersolid state there was no strong motivation
to apply the powerful simulation tools that were under development in those years
to the search for the supersolid phase.

The earlier experimental search for a supersolid phase was focused on the study of 
the plastic flow of objects moving in solid $^4$He\cite{plastic,plastic2}.
Later, the attempts to discover such a state were based on mass flow\cite{greywall,bonfait}
or on TO experiments\cite{bishop}, but no evidence of anomalies
in the behaviour of solid $^4$He was found.
The earlier TO experiment was carried out in the correct range of temperatures (T $>$ 25 mK)
and pressures ($25-48$ bar) but gave no indication of NCRI.
However, under similar conditions,
Kim and Chan were later able to observe NCRI; perhaps the sensitivity of the apparatus was
inadequate or the annealing procedure
reduced the signal to under the detectable level: it still was not time for a revolution.

The only voice out of the chorus has been that of Goodkind and coworkers\cite{goodkind1,goodkind2,goodkind3},
who reported that acoustic waves in solid $^4$He are scattered by a nonphonon family
of excitations with a small activation energy of the order of 1 K\cite{goodkind1}.
Some years later, Ho et al.
observed a sharp peak in the acoustic attenuation in the presence of
$^3$He impurities with a concentration of a few tens per $10^6$,
and the anomaly decreased with increasing $^3$He concentration\cite{goodkind2};
these data were interpreted in a rather unusual way, in terms of
a continuous second-order phase transition from a Bose condensed state
\textit{above} a critical temperature of about $160-180$ mK to a normal state \textit{below} it.
In another more recent experiment\cite{goodkind3}, the interaction between traveling waves
generated by heat pulses and sound waves in pure and impure hcp solid $^4$He was investigated;
the results of this experiment were also interpreted as indicating the presence of 
long-lived nonphonon excitations with
a gap in their spectrum in the high-temperature phase, as previously suggested.
It is worth mentioning these results, even if they are still not understood, because
the dependence on $^3$He concentration and the temperature range in which these phenomena
appear seem to suggest a connection, if not a common origin, with the more
recent quantum effects found in solid $^4$He.

During those years, some attention was devoted to the measurement of the
concentration of not only thermal vacancies, but also hypothetical zero-point vacancies.
A summary of the measurements in solid helium is discussed in ref. \citeonline{simmons}.
It was found difficult to reconcile
the thermal vacancy content measured directly by X-ray diffraction
with values based on the measurements of different quantities
(pressure, nuclear magnetic resonance of $^3$He impurities, ion mobility, plastic deformation, ultrasonic attenuation).
An output of these experimental studies has been the vacancy activation energy; this quantity, measured
by different methods near the melting density of solid $^4$He, was estimated to be in the range of $7-15$ K, with 
the exception of the ultrasonic attenuation measurements of Goodkind\cite{goodkind1}
(who assumed that the nonphonon family of excitations observed in his sound experiments
is directly connected with vacancies), which gave an energy of about 1 K.
In this context, zero-point vacancies have never been experimentally observed.
In principle one can determine the presence of vacancies in a rather straightforward way.
From the wave vectors of the Bragg peaks, one obtains the lattice parameters and the volume
of the unit cell, which can be compared with the value obtained from the macroscopic
density of the system. In reality, this is not so straightforward\cite{simmons3} because one
never has a perfect single crystal completely filling a given volume.
In addition, the direct X-ray method does not have the necessary sensitivity to measure a very small
concentration\cite{simmons3}.
It is estimated that the present experiments cannot exclude ground-state 
vacancies at a concentration lower than about 0.4\%\cite{simmons2}.

During this period,
simulation techniques were used to study liquid and solid $^4$He from
an accurate microscopic point of view, but without trying to answer the supersolid question.
First attempts to obtain a quantitative description of solid $^4$He used variational techniques
and the already introduced JN wave function\cite{hansen}.
Among the variational techniques, it is worth mentioning a new class of wave functions, 
the shadow wave functions (SWFs), which were
introduced in 1988 as a variational tool able to describe liquid and solid phases with the same functional form\cite{vitiello}.
Using an SWF, correlations between particles are introduced implicitly beyond the pair level of $\Psi_J$ by coupling
the particles to a set of subsidiary variables, called shadows, which are integrated over:
\begin{equation}
\Psi_{SWF}(R) = \Psi_J(R) \int_{\cal V} dS \, G(R,S) \Psi^s_J(S)/Q_{SWF}^{1/2} \, ,
\label{eq5}
\end{equation}
where $S=\{{\vec s}_1,..,{\vec s}_{\cal N}\}$ is the many-body configuration
of the shadow variables, $G(R,S)=\prod_{i}^{\cal N} e^{-C|{\vec r}_i-{\vec s}_i|^2}$ is the product
of Gaussian one-body factors, $\Psi_J$ and $\Psi^s_J$ are
Jastrow functions (in general with different pseudopotentials),
which take into account the direct two-body correlations between particles ($\Psi_J$)
and between shadow variables ($\Psi^s_J$), and $Q_{SWF}$ is the normalization
\begin{eqnarray}
Q_{SWF} = \int_{\cal V} dR \int_{\cal V} dS \int_{\cal V} dS' \, \Psi^2_J(R) \times \nonumber \\
\quad \quad \times G(R,S) G(R,S') \Psi^s_J(S) \Psi^s_J(S') \quad .
\label{eq6}
\end{eqnarray}
An SWF is manifestly Bose symmetric, and 
the solid phase is described via a spontaneous broken translational symmetry process;
when the density is above a certain value, the many-body correlations present in the SWF
become so strong that the solid phase is stable and its energy is below that of the metastable liquid phase.
It is important to note that this class of wave functions still provides the most accurate variational description
of liquid and solid $^4$He, accurately predicting the
freezing and melting densities\cite{moroni}; the energy per particle obtained using 
a fully optimized SWF
is only $2-3$\% above that computed using ``exact'' QMC methods and with an almost constant gap
over a wide density range, giving rise to a very accurate equation of state.
Chester's argument was also extended to this class of wave functions;
SWFs thus describe a solid with a finite concentration of vacancies (which has recently been
estimated\cite{arxiv1}) and BEC\cite{masserini}.

``Exact'' simulation techniques provided a more accurate approach to the quantitative microscopic study
of the condensed phases of $^4$He.
Results for the equation of state have been obtained at zero temperature
by the Green's function Monte Carlo method\cite{kalos} (GFMC) and the diffusion Monte Carlo
method\cite{boronatold,moroniold,moroni} (DMC).
These zero-temperature methods are known to rely on the use of accurate variational
trial states to implement the so-called ``importance sampling'', i.e., an improvement of the sampling
algorithm, which restricts the exploration of the configurational space only to the ``relevant'' regions
where the trial state does not vanish, thus substantially reducing the fluctuations in the computed
quantum averages and allowing for their reliable calculation.
This is a delicate aspect of the results obtained by these methods because it not always easy
to show that they are not affected by the choice of the trial state.

This aspect is overcome in another ``exact'' QMC method, the path integral Monte Carlo (PIMC) method
developed by Ceperley and Pollock (for a review see ref. \citeonline{ceperley1}),
which is suitable for computing finite-temperature quantum averages of a many-body system, 
\begin{equation}
\langle \hat{O} \rangle=\textrm{Tr}(\hat{\rho}\hat{O})/\textrm{Tr}(\hat{\rho}) \quad,
\label{eq7}
\end{equation}
where $\hat{\rho}=e^{-\beta\hat{H}}$ is the density matrix.
The method allows for the calculation of
such averages by expressing the {\cal N}-body density matrix
in the coordinate representation at temperature $T=(k_B\beta)^{-1}$ as a path integral:
\begin{equation}
\rho(R,R',\beta)= \int \prod_{i=1}^{M-1}
dR_i \{ \prod_{i=0}^{M-1} \langle R_i | e^{-{\beta \over M}\hat{H}} | R_{i+1} \rangle \} \quad ,
\label{eq8}
\end{equation}
where $R_0\equiv R$ and $R_M\equiv R'$,
and by using an accurate approximation
for the high-temperature density matrices $\langle R_i | e^{-{\beta \over M}\hat{H}} | R_{i+1} \rangle$
such that the quantum averages, computed via the Monte Carlo method,
are not affected by this approximation within the statistical uncertainty of the calculation.
For a boson system the symmetry of $\rho(R,R',\beta)$ is recovered via the direct sampling of permutations
between single-particle paths, which correspond to ring polymers; not only
the superfluid fraction, via the winding-number
technique, but also off-diagonal properties such as the one-body
density matrix, which gives access to the Bose-Einstein condensate fraction, can also be computed\cite{ceperley1}.
Finite-temperature properties have been studied
by the PIMC method\cite{ceperley1}, which
has been used to compute the kinetic energy\cite{ceperley2} in liquid and solid $^4$He
and also the Debye-Waller factor\cite{ceperley3} in solid $^4$He
with good agreement with experimental values deduced from neutron-scattering experiments.

\subsection{Early microscopic studies on a potential supersolid $^4$He phase}
Because of its ability to describe the solid phase via a spontaneous broken translational symmetry process,
the SWF variational technique is particularly suitable to study disorder phenomena in quantum solids.
For instance, the liquid-solid interface was studied\cite{ferrante}
and the first microscopic quantitative estimation of the vacancy activation energy $\Delta E_V$
was another application\cite{pederiva} of the theory.
$\Delta E_V$ was found to depend strongly on the density, and
near the melting density $\Delta E_V$ turned out to be about 15 K;
both of these results were in agreement with the majority of the experimental estimations;
the vacancies were also found to be mobile
since they appeared at many different sites in the lattice during the simulation.

Chester's conjecture on the relationship between vacancies and BEC
was quantitatively verified in 2001 using an SWF in the presence
of vacancies that were found able to induce BEC in the system\cite{galli1}.
The condensate fraction per vacancy, i.e.,
the condensate fraction divided by the concentration of vacancies $n_0^v=n_0/X_v$,
turned out to be about 0.21 $^4$He atoms per vacancy.
In that work, different scenarios for the presence of a BEC phase in solid $^4$He were discussed;
four different phase diagrams were proposed depending on the presence of zero-point vacancies
and on the values of quantities such as the vacancy activation energy and the vacancy effective mass.

Two years later, the same variational technique was used to compute the vacancy excitation spectrum
and longitudinal phonon frequencies in solid $^4$He\cite{galliv,galli3b}.
The calculation was based on the development of the excited-state SWF variational technique
used previously to study the phonon-maxon-roton spectrum in liquid $^4$He with very accurate results.
The agreement of this microscopic and ab initio calculation of the vacancy spectrum
with a tight-binding model was found to be quite
good, and this suggested that a previously introduced
hopping model for a vacancy\cite{heterington,guyer2}
can provide a reasonable approximation when the correct bandwidth is known.
From the computed vacancy excitation spectrum it was possible to extract the vacancy effective mass,
which, in the hcp lattice, turned out to have a small anisotropy and to
be about 0.31 $m_4$ in the basal plane and 0.38 $m_4$
perpendicular to it, where $m_4$ is the mass of one $^4$He atom.
Nevertheless, this value was a large amount of mass for nothing (a hole)!
Also the bandwidth was found to be anisotropic in the range $7-11$ K in hcp
solid $^4$He, depending on the direction of the wave vector in the crystal. 
The large value of the bandwidth $\Delta$ implies
that atoms jump very frequently in the vacant site and that the
residence time of a vacancy, estimated by $\tau=h/\Delta$, is
$0.6\times 10^{-11}$ s, about 4 times larger than the
period of a high-frequency phonon of the crystal.

In 2003 the vacancy activation energy was computed by a projector quantum Monte Carlo method,
the shadow path integral ground state\cite{galli3,galli3c} (SPIGS) method,
developed from the path integral ground state (PIGS)
method\cite{schmidt} which is able to compute exact (within statistical errors) ground state ($T=0$ K)
expectation values.
The aim of the PIGS method is to
obtain a state, $\Psi_{\tau}$, indistinguishable from the true ground state within
the statistical uncertainty of the Monte Carlo calculation by
applying the imaginary time evolution operator
$e^{-\tau \hat{H}}$ to a variational state, which in the SPIGS case is an SWF:
\begin{equation}
\Psi_{\tau} = e^{-\tau \hat{H}} \Psi_{SWF} \quad .
\label{eq9}
\end{equation}
By discretizing the path in imaginary time and exploiting the factorization
property $e^{-(\tau_1+\tau_2) \hat{H}}=e^{-\tau_1 \hat{H}}e^{-\tau_2 \hat{H}}$,
$\Psi_{\tau}$ can be expressed in terms of convolution integrals
that involve the imaginary time propagator $\langle R | e^{-\delta\tau\hat{H}} | R' \rangle$,
which, if $\delta\tau$ is sufficently small, can be approximated as is done with the PIMC method.
This procedure maps the quantum system
into a classical system of special interacting open polymers.
If the starting variational state is Bose symmetric, as in the case of an SWF,
the PIGS method automatically preserves this symmetry; thus, the calculations
obtained by the SPIGS method in the solid phase of $^4$He
provided the first ``exact'' ground-state calculations with the Bose symmetry preserved.
Another appealing feature peculiar to the PIGS method is that in $\Psi_{\tau}$ the variational
ansatz acts only as a starting point, while the full path in
imaginary time is governed by $e^{-\delta\tau \hat{H}}$, which depends only on $\hat{H}$.
Therefore, the role of the starting variational wave function in the PIGS method is drastically
reduced in comparison with other ``exact'' $T=0$ K methods such as GFMC and DMC.
Vacancy activation energies computed by the ``exact'' SPIGS method were found in good 
agreement with the values obtained previously by the SWF technique\cite{galli3,galli3b}.

This was the theoretical and experimental research scenario for solid $^4$He
before the TO ``revolution''.

\section{Recent Experimental Findings: the Revolution Starts and Continues}

In the discussion of more recent experimental results on the investigation of a supersolid phase of $^4$He,
we will discuss experiments not in chronological order but by genre in order to provide
a more complete view of the findings.
We start with the TO experiments.

\subsection{Torsional oscillator experiments}
The scenario was radically changed by the paper by Kim and Chan in 2004\cite{chan1},
in which they reported the measurement of the oscillation period of a TO containing
solid helium confined in Vycor glass,
which is a porous material permeated by interconnected channels with a typical diameter of 60 \AA.
$^4$He inside Vycor glass at low pressure and temperature is a superfluid liquid.
Under compression, $^4$He
solidifies at a pressure larger than that required for a bulk sample. 
Kim and Chan performed the
TO measurement in Vycor glass, and they found that below a temperature of about 0.2 K
the period $\tau_0$ of the oscillator dropped below the value at a higher temperature.
Since $\tau_0=\sqrt{I/G}/2\pi$, where $I$ is the moment of inertia of the TO and $G$ is the torsion
spring constant, if $G$ remains constant this result implies that
the moment of inertia decreased below that corresponding to the rigid rotation of the
system. The fact that there was a critical-velocity effect (at large amplitudes
of the oscillator there was no missing inertia) and that the effect was suppressed
when the $^4$He boson atoms were replaced by 
$^3$He fermions, led to the interpretation of these results as a manifestation 
of NCRI, sought for many years,
associated with the supersolid state in which about 1\% of the $^4$He
atoms did not respond to the imposed oscillation of the container, i.e., the superfluid fraction
$\rho_s/\rho$ was approximately $1\%$.
The critical velocity turned out to be particularly small, of the order of few $\mu$m per second
and of order of the velocity associated with one quantum of vorticity, and also this seemed to fit
in with the supersolid scenario.
These results was announced in 2003 during a talk by Chan at the Quantum Fluids and Solids International Symposium
(QFS2003) in Albuquerque (New Mexico, USA), news that magnetized the interest of many theoretical and
experimental physicists present at the conference. The overall impression was that the confinement
of solid $^4$He was the key ingredient for inducing the necessary degree of disorder in the lattice, probably
in the form of vacancies, to lead the system to a supersolid phase. Only a few months later this 
scenario was cast into serious doubt by a second paper by Kim and Chan\cite{chan2}; the same effect
was measured in bulk solid $^4$He. In this study the authors also noted that
the effect was essentially suppressed by the presence of a plug in an
annular container; the revolution had definitively
started. It seemed that the two measures (Vycor and bulk) should possess a common mechanism at the origin of NCRI.
In 2005 a similar effect was reported by the same group for solid $^4$He confined 
in porous gold\cite{chan3}.
It is important to note that the surface-to-volume ratio available for solid $^4$He 
changes by many orders of magnitude in these experiments; the presence of a similar
phenomenon in the same range of temperatures under such different confinement conditions
is an observation that still needs a
reasonable explanation.
This missing inertia is a robust effect not limited to the lowest pressure region where the solid is stable but 
occurs up to the highest pressures in present experiments of approximately 140 bar\cite{chan4}.
A puzzling result is that the NCRI fraction in bulk $^4$He as function of
pressure has a maximum at a pressure above melting, at about 50 bar\cite{chan4}.
In the light of the wide variation of the NCRI fraction measured later in different high-quality crystal samples
(see below), these results should not be so surprising and might be due to an increased
amount of disorder as pressure is increased.

The experimental results of Kim and Chan have been independently
reproduced by other groups around the world\cite{reppy1,shirahama,kubota,kojima}.
TO experiments have also been replicated under a number of different
conditions in order to extract a greater amount of 
information\cite{kojima,kojima2,reppy2,reppy3,chan4,chan5,chan7,chan8}.
At this point it is important to summarize the findings from the set of 
experimental results obtained by the TO technique.

All the above-mentioned experiments report 
an NCRI fraction, i.e., the superfluid density $\rho_s$ in the supersolid scenario,
which vanishes continuously as the temperature is increased, similarly to a smeared phase transition
rather than a sharp one, as expected in a homogeneous system.
In addition, there is a dissipation peak 
(i.e., a minimum in the oscillation amplitude of the TO), which appears with NCRI
at a temperature where $\rho_s$ has the maximum rate of change.
This is similar to what was measured in TO experiments on $^4$He
adsorbed on planar surfaces\cite{reppy_old}.
The quality of the crystalline state of solid $^4$He strongly depends on the growth
method. Growth by the blocked-capillary method is known to produce polycrystalline samples
of varying sizes and defect contents, depending on the conditions of the growth and the annealing procedure.
Growth at a constant pressure or at constant temperature from the superfluid phase is known to
produce single crystals or almost so. In all but one TO experiments, the solid was grown
by the blocked-capillary method, thus the solid sample was polycrystalline.

Annealing a sample of solid $^4$He is a standard procedure to improve the quality of the crystal,
and the effects of annealing have been reported by Rittner and Reppy\cite{reppy1,reppy2,reppy3}.
They have eben been able to reduce
the NCRI to below the level detectable in their experiment. The influence of annealing on NCRI has been also
observed in other experiments\cite{kubota,chan8};
sometimes the overall effect has been to increase the NCRI fraction,
and sometimes it has reduced it.
However, in contrast with the findings of Rittner and Reppy, it was
never reduced to under the level detectable in the respective experiments.
This indicates that disorder plays an important role in the phenomena.
In fact, a large range of
NCRI fractions (from 0.04\% to about 20\%) have been observed, with
the larger values being observed in samples that presumably contain more defects.
However, the amount of disorder present in the samples was not determined
quantitatively in these experiments. This disorder is likely to be due to grain boundaries,
vacancies, dislocations, or stacking faults.
In one experiment, solid samples grown at constant pressure or temperature were used, and also in these cases
the NCRI signal was detected, although with a reduced amplitude.
On the other hand, high-quality crystals, all exhibiting NCRI, also gave considerably different
values for the NCRI fraction depending on the sample.
As mentioned above,
the NCRI effect also does not set in sharply at a precise ``transition'' temperature, but it continuously
increases up to a constant value with decreasing temperature;
in high-quality crystals\cite{chan5} the transition is much sharper and the NCRI fraction
was found to be in agreement with the expected two-thirds power law for a continuous normal-superfluid
transition (see Fig.1 on J. Phys. Soc. Jpn., Vol.77, No.11, p.111010).
However, as can be seen in Fig.1 (on J. Phys. Soc. Jpn., Vol.77, No.11, p.111010) 
also in the high-quality crystals there is a residual knee
at the temperature where NCRI sets in.
Recently, remarkable hysteretic behaviour has been found in TO experiments\cite{kojima},
with the NCRI phenomenon depending on the history of
the oscillation velocity and on thermal cycling from a temperature above the transition 
temperature to a low temperature.
Another study of the thermal histories of the resonant frequency of 
a TO filled with solid $^4$He resulting from changes in the oscillation speed above
and below the NCRI onset temperature has also been published\cite{chan8}.
It showed that shifts in the resonant frequency 
are irreversibly induced by the variation of the oscillation speed below the onset temperature,
revealing the presence of metastable states;
however, the temperature dependence of the frequency shift for each sample is 
fully reproducible during warming
and cooling scans at low oscillation speeds.
Trapped vorticity might be at the origin of such metastability.

Natural He contains about 0.3 part per million (ppm) $^3$He impurities.
NCRI is extremely sensitive to the concentration of $^3$He.
Additional $^3$He is found to broaden the onset of
NCRI and to shift it to a higher ``transition'' temperature but above a $^3$He concentration of order 0.1\%
NCRI disappears.
More recently ultrapure samples of $^4$He have been studied with concentrations of $^3$He
in the range $1-300$ parts per billion (ppb). At such small concentrations
the effect of $^3$He is also observable on the NCRI signal (see Fig.1 on J. Phys. Soc. Jpn., Vol.77, No.11, p.111010).
A reasonable explanation for this phenomenon is that the $^3$He atoms are trapped by the defects of the crystal.
For instance, there is evidence from ultrasound and shear modulus measurements\cite{dislo1,dislo2}
that a $^3$He atom is bound to the dislocation core by an energy in the range 0.7-3 K.

Rittner and Reppy\cite{reppy3} have also observed an interesting correlation
between the minimum-maximum NCRI measured in
different TO experiments: it increases with increasing surface-to-volume ratio, $S/V$,
which is specific to each TO cell.
This has been interpreted as an indication of the role of disorder in NCRI because a cell with a greater $S/V$,
being more confining, should stabilize more defects in the crystal.

Using a double-resonance TO\cite{kojima,kojima2} Aoki et al. were able to show
that the reduction of the NCRI fraction is characterized by a critical-velocity effect
and not by the displacement amplitude or the acceleration, because in these latter cases the reduction
does not coincide in the two modes;
moreover, they found that the observed dissipations and
frequency shifts of the two modes cannot be made compatible with a model\cite{nussinov} in which NCRI
is explained in terms of the transformation of a liquid component into a glass
component at lower temperatures.
This model also does not explain the different behaviour observed in the TO experiments in the presence of
a blocked annulus.

The value of the NCRI fraction varies by more than 3 orders of magnitude in different samples. Also
the range of temperatures in which NCRI has been detected varies but to a much smaller extent, 
i.e., in the range $60-200$ mK.
When an experiment is conducted on ultrapure high-quality crystal, this range appears to be invariably reduced
to $60-80$ mK.

Since the early measurements were performed in
highly polycrystalline solids, grain boundaries have been suspected to play an important role in NCRI.
The fact that crystals grown at a constant temperature or pressure\cite{chan5}
exhibited NCRI means that grain boundaries are not essential for NCRI,
and other defects must be important.
This also excludes the possibility
that the metastable glassy region of $^4$He, which may be present
in a rapidly quenched sample\cite{grigorev1}, can play the decisive role in these phenomena.
Dislocations are reasonable candidates, since it is known that even in a single
crystal the density of dislocations can vary by a large amount depending on experimental
conditions.
A joint experiment on TO and on the content of dislocations in a single crystal is highly desirable.
This major role of defects in the NCRI of solid $^4$He has even led to the conviction
that the whole issue of the supersolidity of $^4$He is exclusively due to the presence of
extrinsic defects, which any solid sample contains.
By growing solid $^4$He in the mK region, one can obtain single crystals with a very small number of
dislocations\cite{ruutu}, at least dislocations ending on the crystal surface. It will be very important 
to perform TO measurements on such crystals.
In this respect it is important to observe wether high-quality solid $^4$He samples
pass the blocked-annulus test with a reduction of the NCRI fraction compared with that
measured in an open cell, as theoretically expected for a supersolid.

\subsection{Elastic anomalies in solid $^4$He}
Recently, a new anomaly has been found in solid $^4$He\cite{beamish3}.
The elastic shear modulus has been measured and a stiffening of the solid has been 
observed at low temperatures at which NCRI occurs, with an increase of the
shear modulus which is approximately 10\%.
The $T$ dependence of the shear modulus has a remarkable similarity 
with that of NCRI;
also the $^3$He concentration plays a significant role on the shear modulus, and has an effect completely similar
to what it is observed for NCRI in TO experiments (see Fig.2 on J. Phys. Soc. Jpn., Vol.77, No.11, p.111010).
It has been proposed that this anomaly is due to dislocations\cite{beamish3}, i.e.,
the same kind of disorder that has also been discussed in relation to NCRI\cite{chan5,chan8}, 
with dislocations being immobile at 
low temperatures, due to the pinning on $^3$He impurities,
and becoming mobile at higher $T$\cite{beamish3}.
On the other hand, should this completely explain NCRI, it would seem counterintuitive;
the ability to move without dissipation seems at odds with the stiffening of the solid.
The discovery of this stiffening is an important new information on the system, but we are far from
understanding its cause and its relation, if any, with NCRI.

\subsection{Mass flow experiments}
Liquid $^4$He in the superfluid state has a number of peculiar properties in addition to NCRI.
Obviously, in solid $^4$He other signatures of superfluidity have also been looked for.
One method has been to perform experiments in which the measurement of mass flux inside the solid
is attempted.
In the first experiment of this kind, Greywall did not find any indication of superflow upon applying
a small pressure difference between two chambers joined by a capillary and filled
with solid $^4$He\cite{greywall}. Some years later, a second experiment in which a
pressure difference was induced via different liquid-solid interface levels in a U-tube\cite{bonfait}
gave negative results.

The observation of dc flow in solid $^4$He would provide a strong confirmation
of superfluid effects;
therefore, soon after the first positive TO
experiments, searches for such a flow restarted.
In 2005 Day et al.\cite{beamish1} searched for the presence of a pressure-induced flow
of solid $^4$He confined in Vycor glass pores; neither this experiment nor the following one\cite{beamish2},
which was devoted to studying the flow of solid $^4$He through an array of capillaries,
were able to detect any signature of such a flow.
The main characteristic of these two experiments is that the pressure was directly applied
on the solid lattice, thus raising the question of wether the correct ``component'' was being pushed.

In 2006 a different version of the U-tube experiment succeeded in producing observable frictionless
mass flow through the solid\cite{balibar2}; this phenomenon was observed only with the contemporaneous
presence of some grain boundaries directly detected in the cryostat at the liquid-solid interface.
The range of temperatures in which this effect was measured also extends above 1 K suggesting
the possibility that this effect, even if very interesting, could not be related to NCRI 
in TO experiments.
The original interpretation of this result was in terms of the superfluidity of grain boundaries being
the origin of the effect; however, the authors
have recently proposed a different mechanism\cite{balibar3,balibar1}:
the flow of mass along liquid channels present along
the contacts between grain boundaries and walls,
thus regarding the effect as a superflow of the liquid.

Very recently, the results of a new experiment have been announced\cite{hallock}:
Ray and Hallock have observed unusual mass transport in hcp solid $^4$He with an apparatus
that allows the injection of $^4$He atoms from the superfluid directly into the solid using 
two Vycor glass rods, which enter the cell filled with solid $^4$He, and exploiting the fact that $^4$He
inside Vycor glass crystallizes at a much higher pressure than in the bulk.
The flow was observed both in crystals grown by the blocked-capillary method and in crystals
grown at constant temperature.
This flow turns out to be pressure- and temperature-dependent; no flow is observed above 
about 27 bar and above 800 mK.
At low pressure and at $T$ below about 400 mK the flow induced by the unbalanced
pressure between the two Vycor glass rods is approximately linear in time, as expected for
a superfluid moving at a critical velocity.
This characteristic temperature of 400 mK is
at least two times higher than the characteristic NCRI temperature
in the case of low-quality crystals and four times that for high-quality crystals.
In this experiment there seems to be no space for explanations different from a mass flow of $^4$He
atoms inside the bulk solid.
For this reason, should these results be confirmed, 
this experiment may definitively show the presence of superflow in solid $^4$He.
However, given also the different region of characteristic temperature and pressure,
any relation with NCRI seen in
TO experiments is far from being proven and understood,
and further flow measurement at lower temperatures is urgently needed.

\subsection{Thermodynamic properties}
If the establishment of NCRI represents a true phase transition, one should expect to
find some signature of it in equilibrium thermodynamic properties such as, for example, the specific heat.
Early solid $^4$He heat capacity measurements hardly reached a low temperature limit
comparable to the NCRI temperature range; moreover, the sample cells used in those experiments
were constructed with materials (e.g., heavy wall metal) that gave a large contribution to the
heat capacity at low temperature\cite{chan6b}.
This was also the case in the experiment by Clark and Chan\cite{chan_old}, where the authors
extended the specific heat measurement down to 80 mK without finding any sharp features.
More recently, a peak in
the specific heat has been reported at the temperature of the NCRI phenomenon\cite{chan6}
using a sample cell made of silicon, which has a heat capacity more than 10 times lower
than that of solid $^4$He in the considered temperature range (see Fig.3 on J. Phys. Soc. Jpn., Vol.77, No.11, p.111010).
This is a very difficult experiment, and it will be important to obtain an independent
confirmation of this finding. Recently\cite{chan6b}, more accurate measurements have been performed;
they confirm the presence of this peak, and its amplitude depends on the $^3$He concentration
being smaller in ultrapure samples of solid $^4$He (1 ppb). The presence of different concentrations
of $^3$He does not alter the position of the peak, which is around $70-80$ mK.
The specific heat peak also depends on the growth method, with solids grown by 
the blocked-capillary method having a stronger peak.

On the other hand, a very accurate measurement of the melting curve\cite{todoshchenko1,todoshchenko2,todoshchenko3} 
in the temperature range $10-320$ mK
found no contribution to the solid entropy other than the usual contribution due to phonons.
The slope of the melting curve is proportional to the entropy difference between liquid and solid
at coexistence, 
which at very low temperatures should be dominated by phonons.
Therefore, the melting curve gives information on the excess entropic contribution in one of the two phases, and 
it should give sign of a phase transition, being the entropy-variations connected to the specific heat
as function of the temperature.
In the first experiment\cite{todoshchenko1} an anomaly in the melting curve below 80 mK was observed,
but in a later analysis the authors claimed that it was entirely
due to an anomaly in the elastic modulus of Be-Cu,
which their pressure gauge was made of\cite{todoshchenko2}.
They also found a contribution to the entropy proportional to $T^7$ at low temperatures\cite{todoshchenko3}.
Such a contribution is at the limit of their accuracy and
the nature of this term in the entropy of solid $^4$He is still unclear.
Anderson et al., on the basis of a phenomenological theory,
have proposed that such an entropy term may be due to strongly correlated vacancies\cite{anderson1}.
Recently, Maris and Balibar\cite{balibar4} have observed that this term can also be explained
by anharmonic phonon effects.

The absence of an anomaly in the melting line is in serious disagreement with the presence of a peak
in the specific heat since both quantities probe the entropy of the system, and no explanation
for the disagreement has yet been given. Excluding experimental artefacts, one possibility is that the anomaly
in the entropy is absent in the difference between the entropy of the liquid and that of the solid,
since the melting line depends on this difference. If this is the case we do not know its origin.

\subsection{More unexplained anomalies}
Hints of an elusive comprehension of solid $^4$He have also been obtained
from inelastic neutron-scattering experiments.
In 2002, neutron-scattering measurements on bcc solid $^4$He showed the presence of new 
unexplained nonphonon ``optic-like'' excitation modes\cite{polturak1,polturak2},
and in 2006, macroscopic structural fluctuations in bcc solid $^4$He were observed via neutron
diffraction studies\cite{polturak3}.
While a study of the Debye-Waller factor in hcp solid $^4$He
has shown the absence of any low-temperature anomalies down to 140 mK\cite{goodkind4},
recently, anomalous excitations have also been measured by inelastic neutron-scattering experiments
in hcp solid $^4$He\cite{goodkind5}. The authors have identified a branch of
clearly observable excitations with a quadratic dispersion curve centered on a reciprocal lattice 
point: a roton-like mode in solid $^4$He, which was identified as delocalized vacancy modes.
In this respect, a theoretical study\cite{galli4} of ground-state commensurate $^4$He has shown the
presence of vacancy-interstitial pairs as a fluctuation effect.
In variational SWF theory such pairs are unbounded, whereas such pairs turn out to be bounded
in the exact SPIGS method. Since the ground state represents the ``vacuum'' of any excitation
of the system\cite{reattochester}, it has been suggested that such vacancy-interstitial pairs represent
a signature of a new kind of excitation different from phonons.

It is also interesting to remember a study on the free expansion of solid $^4$He
into vacuum\cite{grisenti1,grisenti2}
from a hole of diameter in the range $1-5 \, \mu$m in the container.
First, a sequence of periodic intensity bursts was observed with a very regular period,
which increases with increasing pressure or decreasing temperature,
the so-called \textit{geyser effect}. Above the lambda point at freezing, 1.76 K, the
periodic bursts were observed only on solidification.
At temperatures below 1.76 K, the
pressure at which the geyser effect could be observed was 4-5 bar above the melting curve.
Closer to the melting curve the flow is continuous in time
and pure Bernoulli liquid flow is observed for a pressure of about 2 bar above the melting curve, 
suggesting that the solid closest to the melting curve possesses
different ``flow'' properties\cite{grisenti1}.
Low concentrations of $^3$He, from 1\% down to 0.1\%, are sufficient to
completely remove the ``anomalous'' behaviour near the melting curve,
and the geyser effect is always observable at the melting curve
down to the lowest measured temperature, about 1.3 K\cite{grisenti2}.
In this experiment the system is far from equilibrium, and
excess vacancies were proposed by the authors as a possible mechanism for
the geyser effect, but even if these vacancies could play a role in the geyser effect,
the different flow properties close to the melting line remain unexplained.
We believe that it will be important to extend these experiments to a lower temperature, in the region of NCRI.

\section{Microscopic Studies after the TO ``Revolution''}
An ensemble of $^4$He atoms
represents a strongly interacting many-particle system, which cannot be studied analytically
via a perturbative approach with the aim of obtaining a quantitative microscopic picture
of its physical properties.
The variational method, at least at $T=0$ K, is a viable method but
the boson statistic of the atoms also makes the many-body problem exactly solvable,
at least for an equilibrium state, both at $T=0$ K and at finite $T$, 
by robust simulation methods (quantum Monte Carlo, QMC),
even if one has to be careful because the results refer to a finite number of particles,
usually less than 1000, with periodic boundary conditions.
Therefore, the results of QMC mathods are always affected by finite size effects that should be 
removed before making a comparison 
with experiments. Depending on the quantity, the size effect can be more or less important, and 
it can be estimated by performing computations for systems of different size.
If the system is in an ordered solid state, there is another size effect:
given the periodicity of the crystalline state and the use of periodic boundary
conditions, there is a commensuration effect between the two periodicities,
and this is much more difficult to disentangle.

QMC methods therefore play a leading role in the microscopic theoretical study
of the condensed phases of a strongly interacting many-particle quantum system such as liquid
and solid $^4$He.
QMC methods have been extensively used in the past to acquire valuable
information, especially on the properties of the superfluid phase of $^4$He.
Since the TO ``revolution'', all this expertise has been focused on the solid phase.
A number of studies have also appeared in the last few years
on solid $^4$He based on simplified models and phenomenological
approaches. In the present review we will mainly discuss theoretical results obtained using QMC techniques.

\subsection{Quantum simulations: commensurate solid}
PIMC methods have been applied to study the off-diagonal properties of
solid $^4$He at finite temperatures, and the result of such computations is that a
commensurate solid has a one-body density matrix, which decays exponentially at large distances;
thus, the commensurate crystal does not possess ODLRO\cite{prokofev3,ceperley4}.
Also $\rho_s$ turns out to be zero.
\begin{figure}[tb]
\begin{center}
\includegraphics[width=8 cm]{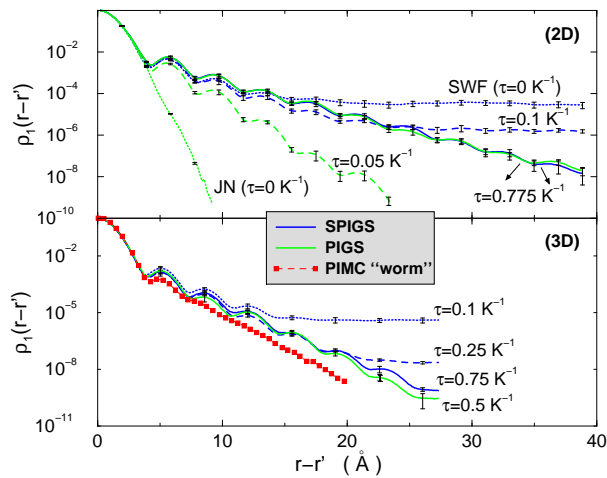}
\end{center}
\caption{(Color online) One-body density matrix computed in commensurate solid $^4$He.
Different colours represent different algorithms used: (red squares) points taken from ref. \citeonline{prokofev1}
obtained by the ``worm algorithm'' PIMC technique, (blue lines) SPIGS method for different $\tau$ values,
(green lines) PIGS method applied to a JN wave function for different $\tau$ values.
(Upper figure) 2D triangular lattice at $\rho=0.0765$ \AA$^{-2}$.
(Lower figure) 3D hcp solid $^4$He near melting.
}
\label{f4}
\end{figure}
The fact that these results, even if not computed at temperatures below 0.2 K, do not
show trends in the range $0.2-1$ K  exclude the possibility of a phase transition at lower temperatures
in the proximity of $T=0.2$ K.
The PIMC method has also 
been used also to calculate exchange frequencies in bulk hcp $^4$He\cite{ceperley5,ceperley5b};
by fitting the computed frequencies to a lattice model
it was concluded that the commensurate crystal is not superfluid.

Even if an SWF describes a quantum system with a finite concentration of vacancies, by choosing
the number of atoms to fit the periodic boundary conditions, one can also use this technique
to simulate a commensurate crystal, and ODLRO has been found with a very small ($5\times 10^{-6}$) 
condensate fraction\cite{galli4}.
This discrepancy between the ``exact'' PIMC method and the variational results has been recently explained
by computing $\rho_1$ by the ``exact'' SPIGS method\cite{galli5}.
The most extensive computations have been performed in a two-dimensional (2D) triangular lattice.
Here the projector $e^{-\tau \hat{H}}$ in eq. (\ref{eq9})
was applied to two completely different wave functions.
The first was an SWF, which is translationally invariant and has BEC, and the second was a JN function,
eq. (\ref{eq3}),
which is not translationally invariant, has no BEC, and is not Bose symmetric.
Note that the application of the Bose symmetric projector $e^{-\tau \hat{H}}$ projects $\Psi_{JN}$
onto the subspace of Bose symmetric wave functions.
As shown in upper Fig. \ref{f4} both computations give the same one-body density matrix $\rho_1$
within statistical errors, and $\rho_1(\vec{r}-\vec{r}')$ turns out to have dominant exponential
decay with increasing $|\vec{r}-\vec{r}'|$ so that there is no BEC.
In the three dimensional case (3D) (see lower Fig. \ref{f4}),
$\rho_1$ obtained starting form SWF and from JN show 
an exponential decay, similar to what was found with PIMC.
Beyond approximately 20 \AA~, $\rho_1$ obtained starting from SWF and from JN slightly differ, and this
suggests that convergence is essentially reached, but larger $\tau$
values are needed in order to obtain complete convergence in this range, a major computational task.
$\rho_1(\vec{r}-\vec{r}')$ in Fig. \ref{f4}, computed by the SPIGS and PIGS 
methods, is shown for $\vec{d}=\vec{r}-\vec{r}'$ 
along the nearest-neighbour direction, whereas $\rho_1$ computed at $T=0.2$ K by the PIMC ``worm algorithm''
is a radial average\cite{prokofev3,prokofev1}, which explains the reduced oscillations superimposed on
the exponential decay.

We conclude that there is overall agreement on the fact that commensurate solid $^4$He does not
possess BEC.
In a very recent computation by the DMC algorithm\cite{boronat}
a vanishing superfluid signal was found, but
an extrapolated estimator for the one-body density matrix gave
a finite condensate fraction for the commensurate solid.
This seems to contradict the result that ODLRO implies superfluidity\cite{leggett1};
perhaps the discrepancy is due to the use of an extrapolated estimator for $\rho_1$.
It was also shown also that a vacancy is able to induce superfluidity, but the effect of correlations between
vacancies has not been studied.

Exact ground-state one-body local densities computed by the SPIGS method has been used to
compute upper bounds for the superfluid fraction using an optimized one-body
phase function\cite{galli_saslow1}. 
The upper bound for commensurate solid $^4$He at melting
was significantly higher (about 20\%) than that experimentally observed,
suggesting an improvement to the theory by
permitting a many-body phase function; a theory for a two-body phase function
has already been developed\cite{galli_saslow2}.

An interesting theoretical problem is why an SWF, a wave function that gives
extremely accurate results for a number of properties of solid $^4$He, gives qualitatively incorrect
results for the long-range limit of $\rho_1$. It has been suggested\cite{galli5} that
the present SWF misses some long-range correlations,
specifically those associated with the zero-point motion of transverse phonons.

\subsection{Quantum simulations: vacancies and other defects}
The calculation of the one-body density matrix, $\rho_1$, in the presence of vacancies has been
obtained by the SPIGS projector method\cite{galli2}.
It was found that a vacancy in a small fcc crystal (107 atoms, 108 lattice sites) with periodic boundary conditions
is able to induce BEC, and that the case of two vacancies was marginally compatible with a condensate fraction
proportional to the vacancy concentration; these results are very similar to those obtained
previously by the SWF technique\cite{galli1}.
However the system was too small to be conclusive, and
multiple vacancies in the hcp crystal were later found by PIMC method
to be strongly correlated\cite{prokofev4,ceperley_vac,prokofev7},
such to form a bound state and phase separation has been proposed\cite{prokofev4}.
Also computations by the SPIGS method\cite{arxiv2} at $T=0$ K show that vacancies are strongly interacting
but they do not form a single compact cluster for up to at least 6 vacancies.
The presence of attractive correlations between vacancies is also partially supported by an
estimate of the elastic interaction potential between two static vacancies in hcp solid $^4$He;
on the basis of an elasticity theory, an attractive interaction has been found
along the basal plane, but the interaction is repulsive in some different directions\cite{mahan}.

Recent calculations of the vacancy effective mass\cite{ceperley_vac,prokofev7} are in reasonable agreement
with the early direct variational SWF calculation\cite{galliv}.
Also the vacancy excitation spectrum\cite{prokofev7} seems comparable\cite{galliv}.

Glass states, grain boundaries, and a screw dislocation have been studied in $^4$He systems
by an implementation of the PIMC method in the grand canonical ensemble\cite{prokofev3,prokofev5,prokofev6}.
The glass states were obtained by equilibrating the simulation at high temperatures
and then ``quenching'' the system down to $T=0.2$ K. This procedure leaves the system in
a metastable amorphous ``solid'' state which possesses ODLRO (up to 0.5\%) and a large
superfluid fraction (up to 0.6). Concluding that the stochastic dynamic of a Monte Carlo sampling algorithm
can resemble the real dynamic of a physical system exceeds the limits of what is reasonable;
however, these results provide a useful insight of what
metastable disorder may be able to induce in solid $^4$He.
Grain boundaries have been found to be generically superfluid, with the exception of highly symmetrical ones;
the transition temperature to the superfluid state has been estimated to be of the order of 0.5 K, thus about
10 times greater than the ``transition temperature'' of NCRI in TO experiments on high-quality
crystals.
Also, the simulation of a screw dislocation has recently revealed the capability of such a defect to sustain
a superfluid flow along its core; however, in order to sustain an NCRI signal of about 1\%,
the authors\cite{prokofev6} have estimated on the basis of the results of their simulation that the required
dislocation density should be about $10^{12}$ cm$^{-2}$, many
orders of magnitude greater than the observed experimental range (see for example ref. \citeonline{disloc}).

Very simplified models of $^4$He confined in porous glass
have been studied by the PIMC method\cite{ceperley6} and the
variational SWF technique\cite{galli6,galli7}.
The PIMC results suggest the presence of a persistent liquid layer just above
the highly localized first layer of $^4$He atoms on the glass wall; this disordered layer
gives rise to a superfluid response. Variational calculations were performed
in a narrow pore with radius  $R=13$ \AA,
and gave evidence that by increasing
the number of atoms in the pore,
solidification takes place layer by layer starting from the pore wall.
Furthermore, the disorder trapped in the system is able to induce BEC even under complete solidification.

The overall scenario of solid $^4$He originating from the microscopic quantum simulation approach
seems compatible with a system which, if commensurate, does not possess the intrinsic property of superflow,
but it is sufficient to include nearly any kind of disorder to obtain a system that can support
some type of superflow.
We can say that the commensurate solid is a marginal ``insulator'', in the sense of the absence of BEC and superflow.
The study of solid $^4$He systems by quantum simulation methods has given a useful insight into
the effect of disorder, but simulation of realistic disorder at zero temperature or at temperatures
of the order of 50 mK will require the study of extended systems and considerable computational resources.

\subsection{Commensurate or incommensurate ground state}
Vacancy activation energies have also been computed by a grand-canonical PIMC
technique\cite{prokofev4}
and more recently by the standard PIMC method\cite{ceperley_vac}.
These calculations substantially agree with previous variational\cite{pederiva}
and exact ground-state calculations\cite{galli3,galli3b}
(the difference of about 3 K at melting is probably due to the different He-He potential employed);
thus, all microscopic QMC methods agree
on the fact that the introduction of an extra vacancy in the ground state
costs a finite amount of energy greater than 10 K.
Moreover, these results do not show any particular dependence
on the system size for up to about 500-1000 $^4$He atoms\cite{pederiva,prokofev4}.
This result of a finite energy cost for a vacancy has been taken as evidence that no vacancies
can be present at low $T$, but
the interpretation of these calculations is far from being
unanimous\cite{prokofev4,prokofev1,anderson1,galli2}.

The nature of the ground state, commensurate or incommensurate, is a very important point
in the discussion on a possible supersolid phase in solid $^4$He. We concluded that the
microscopic calculations on solid $^4$He in literature do not allow one to infer
wether the ground state is commensurate or incommensurate\cite{galli2}.
Obviously this has no more relevance for the
Andreev-Lifshitz-Chester (ALC) scenario\cite{andreev,chester} of essentially
independent zero-point vacancies
because of the strong correlations found between these defects; however, 
it may have relevance for the existence of an incommensurate
ground state with correlated defects; thus, in the following we again propose our considerations.

Even if the ground state has zero-point vacancies, adding one additional vacancy is expected to increase
the energy; this is similar to the case when in superfluid $^4$He one adds one extra phonon to the ground state
that contains the zero-point motion of phonons\cite{reattochester}.
Therefore, we expect that the excitation energy of the system contains a vacancy band with a gap,
but this by itself is not sufficient to exclude an incommensurate ground state.
In addition, in a QMC simulation of a crystalline solid, there is a commensuration effect as mentioned above.
The relevance of such an effect was shown by the following experiment\cite{galli2}.
Take a cubic volume and 108 $^4$He atoms at a density where the solid is stable;
the \textit{exact ground state} is a commensurate fcc crystal.
Now if one places 107 $^4$He atoms in the same volume, the \textit{exact ground state} is incommensurate;
one observes that the number of lattice sites remains 108, one more than the number of particles.

All QMC methods exploit the intrinsic statistical nature of quantum mechanics by
computing quantum averages via the Monte Carlo sampling of a number of configurations 
of the system that have a nonvanishing probability, as measured by the
configurational probability function, which corresponds to the square of the modulus
of a wave function at zero temperature or to the diagonal part of the density matrix 
at a finite temperature.
In order to diminish the commensuration effect one should consider a macroscopic system.
One can argue that all the microscopic approaches that describe the solid phase in a natural way
via a spontaneously broken translational symmetry (Jastrow or SWF for the variational
method, SPIGS or PIMC for the ``exact'' methods) have a configurational probability function
that gives a nonzero probability, not only for configurations that correspond to a commensurate
crystal but also for those that correspond to an incommensurate crystal.
This is due to the fact that a vanishing
probability is only obtained for configurations with two or more particles
at a distance lower than the hard-core radius of the interaction potential.
Defects such as
vacancies, interstitials, and also grain boundaries, dislocations and more complex defects
in the solid phase do not give rise to such configurations with overlapping atoms,
so that such defects have a nonvanishing probability of being present in the ground state.

Thus, configurations corresponding to a commensurate crystal
as well as configurations corresponding to an incommensurate crystal
will contribute to the normalization of a Jastrow wave function $Q_J$ (eq. (\ref{eq2})),
of a shadow wave function $Q_{SWF}$ (eq. (\ref{eq6}))
\textit{but also} to the normalization of the SPIGS
wave function $\Psi_{\tau}$, and to the bosonic partition function of $^4$He in the PIMC
\begin{equation}
Z_B={1\over \cal{N}!}\sum_{\cal P} \int dR \, \rho(R,{\cal P}R,\beta)
\label{eq10}
\end{equation}
(${\cal P}$ in eq. (\ref{eq10})
represents the sum over all possible $\cal{N}!$ permutations imposed by the Bose statistic).
The point is: which is the region of the configurational space which gives the dominant
contribution to the normalization for the macroscopic system we are considering?
For $Q_J$ and $Q_{SWF}$, Chester's argument proves that the overwhelming contribution
originates from regions of the configurational space where the crystal is incommensurate, and
this is also true for $Z_B$ in eq. (\ref{eq10})
for any finite $T$.
Recently, we have computed the equilibrium concentration of vacancies $X_v$ in the ground state of solid
$^4$He described in terms of an SWF\cite{arxiv1}.
Close to the melting density we obtained $X_v=1.4\times 10^{-3}$,
a value much larger than that found for the Jastrow case.
Our value is below the present experimental upper bound, but it is quantitatively significant.
For instance, should the vacancies organize themselves in dislocations,
one would find a dislocation density of
$n_D\simeq 10^{10}$ cm$^{-2}$ when $X_v \simeq 10^{-3}$.
The value of $X_v$ obtained from $Q_{SWF}$ is rather large but not unreasonable;
however, one should be cautious. First,
using the SWF, vacancies are not strongly interacting\cite{arxiv1} as in exact computations.
Second, we do not know what happens to $X_v$ given by the SPIGS $\Psi_{\tau}$ in the limit
$\tau\to\infty$.
Similarly we do not know what happens to $Z_B$ in eq. (\ref{eq10}),
in the $T\to0$ limit.
We conclude that at present the question of whether the ground state of solid $^4$He
is commensurate or incommensurate is still unanswered
and that answering this question is a priority.

\subsection{Phenomenological theories}
In two recent papers Anderson\cite{anderson2,anderson3}
suggested that the phenomenon observed up to now is not supersolidity, which might occur at a
lower temperature, but a dissipative regime dominated by quantized vortices.
An analogy is made with high-temperature superconductivity above the superconductor
phase transition in a regime where a Cooper-pair gap is still present even if the phase fluctuations destroy
standard superfluidity. In this regime the system is better described
in terms of an incompressible vortex fluid, which manifests itself, for instance,
with an anomalous diamagnetic susceptibility, that is the quantity for a charged system which is the analogous of 
NCRI in the neutral system.
Some aspects of NCRI in solid $^4$He, such as the critical velocity
and some observed metastability phenomena, are consistent with such a vortex fluid view of this state.
In this picture, the system should be considered\cite{anderson2} as a density wave rather than a solid,
with the helium atoms flowing as a fluid through this density wave.

On the basis of a phenomenological Landau theory,
with the assumption that the hypothetical supersolid phase transition is continuous,
Dorsey et al.\cite{dorsey} studied the transition in $^4$He by assuming that superfluidity is coupled 
with the elasticity
of the crystalline lattice. They found that this assumption does not affect the universal properties
of the superfluid transition and that a lambda-anomaly in the specific heat should appear.
They also predicted anomalies in the elastic moduli near the transition and local variations
in the superfluid transition temperature due to inhomogeneous lattice strains.

Toner has recently used Landau theory to study the effect of quenched edge dislocations on a possible supersolid
transition\cite{toner}. He found that quenched dislocations can promote the transition from a normal solid
to a supersolid even if the dislocation-free crystal is normal down to $T=0$ K.

On the basis of a Ginzburg-Landau theory, Ye\cite{ye1}
has studied the possible phase transitions in $^4$He and the conditions for the existence of
a supersolid phase; more recently he has investigated
the elementary low-energy excitations inside a supersolid\cite{ye2}.

The quantum mean field approximation for bosons at $T=0$ K leads to the Gross-Pitaevskii (GP) equation,
a nonlinear Schr\"odinger equation. This equation is very successful in describing low-density cold atoms but has
limited application to the strongly interacting superfluid $^4$He. Pomeau and Rica\cite{pomeau1}
observed in 1994 that the GP equation with a nonlocal interaction kernel produces a crystalline
state when the density is sufficently large. The dynamical properties of the model\cite{pomeau2,pomeau3}
display NCRI and a solidlike elastic response to a small stress. The nonlocality of the interatomic 
interaction has a key role in the model, but no account is given of the short-range correlations
arising from the repulsive short-range interatomic interaction; thus, the results are interesting but it is
unclear how relevant they are for solid $^4$He.

\subsection{Discussion: present understanding}
After the large number of experimental results published in the last four years,
one can have confidence in the quantum nature of the anomalies
measured in solid $^4$He.
However, what is the origin of the anomalies in solid $^4$He?
Are they intrinsic or extrinsic? Also, what is the nature of the ground state of solid $^4$He?
Is it commensurate or incommensurate?
The combination of the possible answers to these questions gives four
possibilities; the fifth is the vortex liquid suggested by Anderson, but
if the experiment by Hallock is confirmed, this case can probably be ruled out.
Can we exclude any of the other possibilities?
The case of an incommensurate ground state that is supersolid only due to extrinsic
disorder does not seem reasonable and we will not discuss it.

We first consider, for solid $^4$He, a commensurate ground state that possesses
BEC or superflow as an intrinsic property.
As we have previously discussed, exact
microscopic simulations have convincingly shown that given the He-He interaction potential,
a commensurate solid does not possess ODLRO; thus, we can exclude the intrinsic-commensurate case.
Moreover, in refs. \citeonline{prokofev2} and \citeonline{prokofev1} the authors propose a series
of considerations, specific to solids in which the translational symmetry is spontaneously broken,
that support the conclusion that a commensurate supersolid has zero probability
of being found in nature because it requires an accidental symmetry between
vacancies and interstitials, which are the individuated necessary ingredients for a quantum solid
to show superfluidity\cite{prokofev2}.
We find such considerations vague and not very convincing.
However, if we accept them,
the ODLRO found in a commensurate solid described using an SWF\cite{galli4} is consistent with
this argument; in fact, the SWF state, which is the exact ground state of an unknown Hamiltonian (not that of 
a system of $^4$He atoms), has a finite concentration of vacancies, and
its simulation in a commensurate
state is equivalent to forcing it to possess the previously discussed accidental symmetry,
and, in fact, unbound vacancy-interstitial pairs are present in the SWF\cite{galli4}.
Also, the recently introduced symmetric JN wave function\cite{boronat} cannot be considered
a counterexample; this state contains one-body terms that explicitly break the
translational symmetry, and thus represents
the ground state of a Hamiltonian that is not translationally invariant.
If the above-mentioned considerations are true, a system whose ground state corresponds to a
commensurate solid cannot be supersolid; no arguments or counterexamples against this conclusion
are known.

We now consider the intrinsic-incommensurate case.
This case not only includes the ALC scenario\cite{andreev,chester} of zero-point vacancies,
but also that in which the ground state contains a correlated state
of vacancies\cite{anderson1} or vacancies and interstitials\cite{dai}. One can also consider 
other defects\cite{toner} or a correlated mixture of different defects.
There is a consensus on the fact that the 
ALC scenario\cite{andreev,chester} of essentially independent vacancies
cannot alone be the 
explanation of the NCRI and of other anomalies recently observed in solid $^4$He.
Even if we accept that the simple estimation of the vacancy activation energy, giving a positive energy cost
in systems with less than a
few thousand atoms, does not exclude the presence of ground-state vacancies\cite{galli2},
the observation of strong correlations between vacancies with some
indication of a possible bound state\cite{prokofev4,ceperley_vac,prokofev7,arxiv2}
leaves only the possibility of a scenario in which, if the ground state contains vacancies, such
defects are highly correlated\cite{dai,anderson1}. 
One can argue that this is not probable but there is no strong evidence to exclude it.
The case of correlated vacancies
can be ruled out if phase separation occurs; our personal view is that
present computations are not conclusive regarding the issue of phase separation, and studies of larger
systems with increasing number of vacancies are needed.
It is possible, for example, that vacancies are converted into other defects or that the continuous
mutation into different defects occurs recursively.
However, we should also consider the possibility that a hypothetical supersolid phase might be
associated not with vacancies but with some other correlated defects that, in any case,
would make solid $^4$He different from the textbook picture in which each atom occupies
a crystal cell. We know that the solid is periodic; this is demonstrated by
the observation of Bragg peaks in the scattering of neutrons or X-rays, but how
the atoms populate the crystal cells may offer some surprise.
The fact that we still do not have sufficient elements to exclude the incommensurate-correlated-defect
ground state also indicates that we cannot use the microscopic results of quantum simulations,
in which no ODLRO has been found for the commensurate case, to deduce that the ground state
of solid $^4$He is commensurate and that the conditions of these simulations were not an accidental 
realization of the system.

We now consider the remaining possibility, the extrinsic-commensurate case,
i.e., the possibility that all the ``supersolid'' phenomena in solid $^4$He (with a commensurate ground state)
are due to the presence of some disorder
induced by external conditions such as the condition of crystal growth. If this were
the case, it would be of interest per se but, of course, not as interesting as if
supersolidity were an intrinsic property. In such a case if a more and more
perfect crystal is grown, NCRI would disappear and no effect will remain,
leaving the ground state of solid $^4$He qualitatively similar to that of a normal solid.
Microscopic quantum simulations have shown that almost any kind of disorder is able to
induce superfluidity in the system; however, on the basis of the same results,
no defect has been found to be capable of clarifying the 
observed anomalies.

As mentioned above, the evidence from exact simulations is that 
commensurate solid $^4$He has no ODLRO
and no superfluidity.
An interesting question is whether this result depends on the form of the interatomic interaction.
It is possible that a commensurate solid might have ODLRO and superfluidity if the interatomic interaction 
differs substantially from that appropriate for $^4$He.

\section{Conclusions}
The recent surge of interest in solid $^4$He has provided much new information on the properties of this
fascinating solid. Anomalies now are not limited to TO evidence for NCRI but are also present
in the shear modulus and specific heat.
Solid $^4$He has turned out to be much more difficult and interesting
than expected, and we do not yet have a clear overall picture of the properties of this 
system. All attempts to explain the behaviour of solid $^4$He in more conventional terms 
have failed so far, and the presence of a nonconventional state appears to be the 
most likely explanation.
Should the report of superflow in solid $^4$He be confirmed, this conclusion will be definitive.
However, the nature of this nonconventional state is still unclear. 
Experiments as well as microscopic theories indicates that the TO results, even if they are due to supersolidity,
cannot be interpreted in terms of the original ALC scenario of weakly interacting vacancies.
Vacancies are strongly interacting and experiments clearly indicate that other defects,
such as dislocations, grain boundaries, and $^3$He impurities, must have an important role.
Actually, the measured quantities depend so sensitively on the growth method, annealing, and
$^3$He content that some researchers believe that all the anomalies are only due to extrinsic defects.
The question of whether such a nonconventional state is indeed only due to extrinsic defects or whether
it is an intrinsic property of this quantum solid is a fundamental question that requires
an answer. An argument in favour of the intrinsic nature of the state is as follows. 
Although, as already mentioned, the value of the NCRI fraction varies by many orders of magnitude 
depending on the experimental details, the transition temperature is affected only marginally
by such details.
If supersolidity were entirely due to extrinsic defects, 
it would be rather difficult to understand why the transition temperature is 
hardly affected when the number of defects varies by many orders of magnitude. 
In fact, the presence of quantum effects also in high-quality $^4$He solid samples
points in the direction of a nonconventional state of strongly correlated-defects;
a model of this kind, based on a network of dislocations was developed in ref. \citeonline{toner},
and the transition temperature turned out to depend strongly on the defect concentration.
Obviously, this argument does not exclude the possibility that some other
extrinsic-defect model might agree with experiments; we simply
observe that a more natural explanation comes from the other possibility as follows.
If the underlying ground state has BEC, 
the system will have its own 
transition temperature but the intrinsic value of the NCRI fraction may be very small. 
The presence of extrinsic defects might have an amplifying effect on the NCRI fraction without 
necessarily markedly affecting the transition temperature, i.e., extrinsic defects might 
change the NCRI fraction without significantly affecting other properties of the system.
The present theoretical understanding is that a commensurate crystal is normal even at zero temperature and
that this commensurate normal state is marginally stable, i.e., almost any deviation from it 
leads to a superfluid response.
Therefore, if the anomalies of solid $^4$He below a temperature in the range $70-200$ mK
are due to intrinsic supersolidity modified by the presence of extrinsic disorder, $^4$He in the ground state
ground state cannot be a commensurate crystal.
Some researchers argue against this possibility because,
according to present computations there is an energy cost upon
introducing any of the defects considered so far.
Others researcher argue that this is not convincing because one is neglecting the 
commensuration effects present in quantum simulations.
In our opinion the question of whether the ground state is commensurate or incommensurate is still open
as a fundamental question, and hopefully experiments with improved sensitivity
will give an answer to whether intrinsic disorder exists even at the lowest temperatures.
From theory,
we also expect extended investigations in a more realistic way of defects
like multiple vacancies or dislocations as well as the study of the microscopic dynamics
of the system.

From the experimental point of view, it is important to perform joint measurements
on the same sample using different techniques in order to characterize solid $^4$He
more completely. The expertise of growing crystalline $^4$He of the highest quality
should be used to investigate supersolidity phenomena.
These are just a few examples of studies that should be undertaken.
The issue of supersolidity will probably remain at the centre of attention
in future. We have learnt something but much more has to come.

Note added in proof:
During the 2008 summer new experimental results have been announced;
in the following, we report on some of them.
Rittner and Reppy have performed an enhanced version of the blocked-annulus TO experiment which
confirm that NCRI effects are due to inertia and not to stiffness variations of the solid sample
(A. S. C. Rittner, workshop "Supersolid 2008", 18-22 August 2008, ICTP, Trieste, Italy).
As reported by Beamish (J. Beamish, workshop "Supersolid 2008", 18-22 August 2008, ICTP, Trieste, Italy),
the shear modulus anomaly is not at the origin of NCRI and is not an effect of statistics because the anomaly
is present also in hcp $^3$He and it is essentially the same as in hcp $^4$He.
On the other hand, no anomaly is present 
in bcc $^3$He. TO experiments (group of M. H. W. Chan) on solid $^3$He samples,
grown from the same helium gases used by Beamish, show no NCRI signal from either hcp and bcc solid $^3$He.
Also, more accurate specific heat measurements in solid $^4$He with different $^3$He concentrations
indicate that the magnitude of the specific heat peak does not depend on the $^3$He concentration
but it does depend on the crystal quality
(M. H. W. Chan, ULT 2008, 14-17 August 2008, Royal Holloway University of London, London, UK).

\section*{Acknowledgments}
D. E. Galli would like to acknowledge useful discussions with N. V. Prokof'ev and M. H. W. Chan.

He would also like to thank Eugenia, Giacomo, Matteo, and Benedetta for their patience.

\end{document}